# Anomaly-Based Intrusion Detection by Machine Learning: A Case Study on Probing Attacks to an Institutional Network

**EMRAH TUFAN[1], CİHANGİR TEZCAN[1], AND CENGİZ ACARTÜRK[1,2], (Member, IEEE)**
[1]Department of Cyber Security, Middle East Technical University, 06800 Ankara, Turkey
[2]Department of Cognitive Science, Middle East Technical University, 06800 Ankara, Turkey

Corresponding author: Cengiz Acartürk (acarturk@metu.edu.tr)

**ABSTRACT** Cyber attacks constitute a significant threat to organizations with implications ranging from economic, reputational, and legal consequences. As cybercriminals' techniques get sophisticated, information security professionals face a more significant challenge to protecting information systems. In today's interconnected realm of computer systems, each attack vector has a network dimension. The present study investigates network intrusion attempts with anomaly-based machine learning models to provide better protection than the conventional misuse-based models. Two models, namely an ensemble learning model and a convolutional neural network model, were built and implemented on a data set gathered from a real-life, institutional production environment. To demonstrate the models' reliability and validity, they were applied to the UNSW-NB15 benchmarking data set. The type of attack was limited to probing attacks to keep the scope of the study manageable. The findings revealed high accuracy rates, the CNN model being slightly more accurate.

**INDEX TERMS** Anomaly-based, misuse-based, intrusion detection systems, probing attacks.

## I. INTRODUCTION

Humans are more dependent on ICT (Internet and Communication Technologies) than ever, and this trend rapidly grows with time. Moreover, one characteristic that makes ICT a game-changer technology is computers' ability to inter-communicate in a network. Attempting to exploit this capability, criminal hackers target governmental or non-governmental organizations to cause financial, political, physical, and psychological damage. The increasing importance of cybersecurity threats accompanies this situation.

Numerous announcements are made periodically to emphasize the prevalence of risk exposed by cybercriminals. Cisco reports preventing six trillion threats in 2018, more than 23,000 threats per second [1]. The reported global financial damage caused by cyber-attacks cost 522 billion dollars in 2018, and this amount was nearly doubled with 945 billion in 2020 [2]. According to Check Point's report, top six attack types in 2019 include crypto miners (38%), botnet (28%), mobile (27%), banking (18%), infostealer (18) and ransomware (7%) [3].

European Agency for Network and Information Security (ENISA) announced top fifteen attack types in 2019-2020 [4]. All declared categories more or less involve network component as a part of attack vector. More interestingly, ENISA's report emphasized the increasing popularity of large organizations such as governments and corporations as targets among attackers.

Growing trend and immensity of the threat against enterprise IT infrastructure motivated the authors to do research on this field. Network security plays a primary role in alleviating the cybersecurity risks in an organization. Accordingly, for businesses and organizations, detecting and preventing cyber-attacks are of significant importance today. Besides other components in defending side's arsenal, organizations need robust intrusion detection systems (IDS) to protect information systems.

Conventional, rule-based intrusion detection systems are fast, and they can operate on low computing resources. So, rule-based IDS is usually appropriate for an environment with many sessions per second in heavy network traffic.

The associate editor coordinating the review of this manuscript and approving it for publication was Rosalia Maglietta.

On the other hand, conventional IDS mechanisms are limited in their flexibility, thus being disadvantageous in detecting novel types of attacks. This article investigates whether machine learning (ML) models enhance intrusion detection by reducing false detection rates in a real-life environment. We developed two ML models and trained the models by the data obtained from an institutional network. We have also focused on a specific type of attack, namely probing, to keep the scope of the investigation manageable.

There exist similar studies which have the main focus on IDS with both rule-based and anomaly-based approach. However, most of the available studies utilize several common public data sets, as presented in detail in Section 3. Public data sets are valuable resources for benchmarking and educative purposes. However, they are limited in reflecting the enterprise environment targeted by cybercriminals. This article's major contribution is that it gathered the data from an institutional production environment and built the ML models on top of it. Therefore, we discuss employing ML learning methods in a natural setting for protecting an enterprise network.

The article is organized as follows. In Section 2, the background on IDS and ML concepts and a brief review of the relevant work are presented. Section 3 introduces the methodology of the investigation. Section 4 reports performance evaluation criteria and the results. Finally, Sections 5 and 6 present a discussion of the findings, the conclusion, and future work.

## II. BACKGROUND AND RELATED WORK
### A. INTRUSION DETECTION SYSTEMS

Intrusion detection is a well-known problem in cybersecurity research and practice. Numerous approaches have been proposed to develop better intrusion detection systems (IDS). Figure 1 presents a general taxonomy of IDS mechanisms based on data collection methods and attack detecting techniques [5].

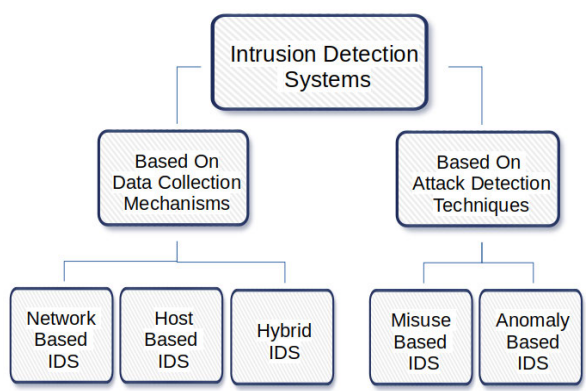

**FIGURE 1.** Intrusion Detection Systems taxonomy based on data collection methods and attack detecting techniques [4].

Among the IDS mechanisms based on data collection, the host-based IDS (HIDS) mechanisms analyze behavior at the endpoints, namely the hosts. HIDS mechanism is based on monitoring system resources, such as memory consumption, CPU utilization, disk I/O, and tracking the processes to detect any abnormal activity. So, the scope of the responsibility of a HIDS is the host on which it operates [6]. In contrast to a HIDS, a network-based IDS (NIDS) captures the data transferred through the network by employing sensors and subsequently analyzing them to detect suspicious incidents [7]. A NIDS may work standalone at the entry point of the protected network to capture and inspect each incoming/outgoing packet, or it may be placed in another network segment. The traffic may be sent to a NIDS for inspection by employing tap devices planted at critical nodes. As its name suggests, a hybrid IDS aims at combining the strengths of both HIDS and NIDS. This approach advocates the IDS placement both on hosts to inspect local incidents and on the network to analyze the traffic flow. Although the idea sounds plausible, Hybrid IDS mechanisms' primary challenge is the high administrative load for information security professionals.

A classification of IDS based on attack detecting techniques reveals two groups: Misuse-based IDS mechanisms target known attacks by detecting the specific behavior exhibited by an attack, in other words, its signature [8]. Misuse-based systems constitute an effective solution against intrusion attempts that are well-recognized and documented. However, misuse-based IDS mechanisms have disadvantages. They may not detect unknown or zero-day attacks. Moreover, each attack type needs to be converted to a rule to ensure detection, requiring frequent and timely updates for adequate protection.

In contrast to misuse-based IDS, an anomaly-based IDS works on the principle of detecting a deviation from the expected system behavior. For instance, common anomaly-based network IDS mechanisms usually observe irregular congestion or unexpected behavior occurrences, such as too many packet re-transmissions. Accordingly, an anomaly-based IDS' overarching goal is to analyze and categorize network flow to discriminate malicious attempts from everyday occurrences [8]. In contrast to the misuse-based approach, there are no specific rules to classify what constitutes an intrusion attempt.

The core of an anomaly-based IDS mechanism is the learning component. The following section introduces ML techniques for anomaly detection within the IDS design and implementation framework.

### B. MACHINE LEARNING TECHNIQUES FOR ANOMALY DETECTION

Recently, there exist numerous approaches for the classification of machine learning methods for IDS design. A broad taxonomy suggests five ML methods for anomaly-based intrusion detection, as shown in Figure 2 [9]. Basically, within the framework of IDS mechanisms, supervised learning approaches assume the presence of a distinction in the data set, which labels a flow/packet's status as an attack or not. Parametric supervised models summarize the data with a fixed set of parameters independent of the sample size, whereas in non-parametric models, the parameters do exist, but they are subject to change depending on the size of the data set [10]. Methods such as logistic regression and naive Bayes are classified as parametric models, whereas k-nearest neighbors (KNN), decision trees, and support vector machines (SVM) are classified as non-parametric methods. For example, implementing the supervised learning methods for IDS, Chikrakar and Chuanhe [11] proposed a model in which KNN grouped similar data samples, output clusters were classified by SVM classifiers with high accuracy values. The major challenge in those systems is the models' limited scalability in that those models may not be robust against large data sets.

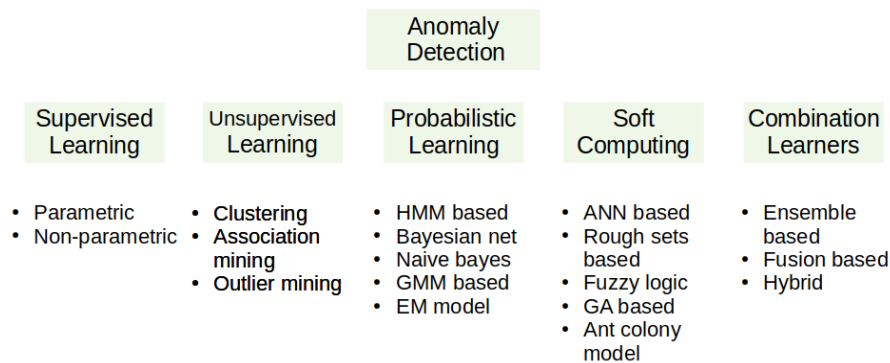

**FIGURE 2.** Machine learning methods for anomaly-based intrusion detection.

Unsupervised learning has the advantage of being able to be built without labeled data sets. Given that the labeling process in supervised learning often requires human expertise and manual annotation in some cases, this is a significant advantage. Among specific types of unsupervised learning models, clustering methods do neither require classes nor labels. A clustering model discriminates between clusters by evaluating the given training data set. Association mining aims at identifying events that occur together frequently. It utilizes statistical methods to make inferences for future instances. Outlier mining looks for patterns that do not match the majority of data. Syarif et al. [12] proposed a model using five different clustering algorithms and compared results with misuse-based models. Four of their models performed better than their counterparts. However, high false-positive rates were acknowledged as an area in need of further development.

Probabilistic learning models estimate new instances affected by randomness and other probabilistic uncertainty [9]. A probabilistic model's distinctive feature is updating the previous estimates based on new evidence learned from training data. Probabilistic learning has been used for IDS design. For example, Aung and Oo's [13] frequent pattern growth algorithm used item sets to build frequent pattern trees, which are used to detect anomalies in the traffic.

Soft computing models foreground the utilization of the randomness factor, in contrast to conventional, exact communication in computing systems. The problem may be resolved by using probabilistic models and fuzzy sets. As an example application of the approach to IDS design, Hamamoto et al. [14] proposed a model that combined fuzzy logic and genetic algorithm. The former was used for generating dynamic signatures from network flow, and the latter classified the instances as anomalous or not. The authors claimed a high detection rate with a low false-positive rate of below 1%, remarkably accurate in anomaly-based IDS. A disadvantage of the model is its high complexity stemming from the combination of numerous methods.

Combination learners, as the name implies, combine multiple learning techniques. The goal is to alleviate each model's weaknesses and increase the overall capability of the combined model. Ensemble methods, fusion methods, and hybrid methods can be placed under this category. Pham et al. [15] proposed a model for IDS, which employed an ensemble method. This model compared two ensemble methods, namely bagging and boosting. J48 decision tree classifier was used as the base classifier. The models were tested on the Network Security Laboratory - Knowledge Data Discovery (NSL-KDD) data set. The results revealed that the bagging method was more accurate in both detection and false-positive rates.

In summary, a review of the previous work on the use of ML models for IDS design reveals that there are driving factors that encourage researchers to employ a specific ML method. Those factors include the availability of labeled data, the proposed system logic in real-time or offline, the computation power available to use, the presence of intervention by field experts, and the volume of the data to be processed for a second.

As for the design and development of anomaly-based IDS mechanisms, there is no single ML method over the others. This situation indicates the need for further investigations of the use of ML methods for IDS design, which address specific attack vectors and their context of use (i.e., the presence of current data recorded in real-life settings vs. the available data sets for research purposes). The present study presents an evaluation of two ML models developed for this study, as presented in the following section.

## III. METHODOLOGY
### A. THE DATASETS

Machine Learning models need data for training; anomaly-based IDS are no exception. There are mainly two categories of data sets, namely public data sets and private data sets. Researchers usually use public data sets for benchmarking purposes.

A frequently used public data set for network security analysis is provided by the Defense Advanced Research Projects Agency (DARPA). The *DARPA 1998-1999 data set* was created in a lab environment with artificially designed cyber attacks. The data include application protocol traffic, such as FTP, Telnet, IRC, and SNMP. The attack types are probing attacks, rootkit, buffer overflow, and DoS (Denial of Service). The main issue about the DARPA 1998-1999 data set is that it is outdated, so being limited in terms of coverage of more recent attacks and the network architecture of the lab environment [16]. Another popular public data set is the *Knowledge Discovery and Data Mining (KDD) 1999 data set*, an updated version and the first derivative of the DARPA 1998-1999 data set. It has broader coverage in terms of the attack instances than its parent (e.g., *smurf* attacks and *neptune* attacks). It covers four classes of attacks, namely *DoS*, *user to root*, *remote to local* and *probing* attacks. Since the KDD 1999 data set was partially redundant and included corrupt data, a follow-up data set; namely, the NSL-KDD was built. Nevertheless, NSL-KDD is also an outdated data set, lacking recent attack types, like the DARPA 1998-1999 data set [17].

More recent public data sets exist for IDS design and development, such as the UNSW-NB15 data set [18]. It was created in a lab environment with artificially designed malicious and benign instances by a commercial network simulation device. Other example data sets for IDS research include CICIDS data sets by the Canadian Institute for Cybersecurity, such as the CICIDS2017 [19]. The CICIDS data sets were also created in a lab environment.

Those data sets have been providing valuable resources for the researchers. The majority of the previous work has employed public or private data sets prepared within lab environments with network traffic simulators (henceforth, *simulated data*). Despite the presence of exceptions that employed real-world data in IDS research, e.g., [9], more research is needed that employs data obtained from daily settings. In particular, simulated data sets approximate the users' behavior profiling in a specific enterprise network rather than representing the user behavior veridically. Therefore, they are relatively limited in providing a comprehensive representation of threats in daily settings.

Another limitation of the simulated data sets is their coverage. The issue is not about the limited scope of various attack types. On the contrary, the richness of the attack variety is a barrier against an efficient representation of a specific network and a factor that may reduce ML model training's efficiency. Specific enterprise networks have their peculiar services and network architectures. Thus, if one trains a model with a general-purpose, simulated data set and puts this model on their enterprise IDS, this model would perform under its potential. In real-life environments, the malicious simulated traffic and regular traffic reflect peculiar enterprise user behavior. To sum up, enterprise IDS models need to be trained with real-world traffic to provide more reliability.

A further limitation of the simulated data sets is that attack simulators try their best to create attacks per discovered vulnerabilities. However, different payloads may be created in a real-life scenario that exploits the same weakness. For instance, the Hulk is a well-known and recent example of application layer DoS attacks. Despite the availability of its most widespread implementation by Shteiman's python script [20], numerous derivatives of the original script are available in public repositories. An IDS model would be limited in scope if trained under ideal conditions provided by attack simulators. It is possible to expand the coverage of the attack simulator.

Apart from the limitations, there are advantages which render simulated data sets more tempting and therefore widely used. For instance, since the attacks on them are controlled, they are precisely labelled unlike their real-world counterparts. Moreover, they are generic and able to reflect an average environment with commonly used services. On the other hand, real life data sets are organization specific and may not be generalized. And lastly, simulated data sets are mostly public and easily available for everyone.

Nevertheless, it will stay as an approximation to the real-world situation rather than veridically represent it. A novel aspect of the present study is that we tested the ML models by collecting data from a real-world environment by taking necessary institutional permissions. We call this private data set the *institutional data set*, presented below.

#### 1) THE INSTITUTIONAL DATA SET

The data were collected from the pop-up network interface of the organization with `tcpdump`.[1] The sessions that were initiated from outside sources were recorded. Figure 3 demonstrates the simplified setting where the data were collected.

In Figure 3, the traffic captures were taken from network interface A. In total, five captures were taken in daytime/working hours and weekends or after working hours. The flows/sessions were extracted from the packets, then shuffled and randomly sampled to produce a representative and manageable data set. The final data set included 100,000 flows/sessions, after around 481,000 sessions before the random sampling.

The sessions were initiated by 38,804 unique hosts. Although this number seems high, most of the traffic came from a limited number of clients. The top 20 of them initiated 36% of all the sessions. Similar behavior was observed in the destinations such that the top 5 received 57% of the whole sessions. The destination port and the requested services demonstrated heterogeneous distribution, as well. Independent of their availability, *HTTP*, *DNS*, *telnet*, *SMB*, *SQL*, *SMTP*, and *SSH* were the most demanded data set applications. Their weight reached 55% of all sessions. The average duration of a session was 6.06 seconds, the average size of the exchanges was 41,385 bytes, and the average packet count was 62 for a single session.

---
[1]https://www.tcpdump.org/

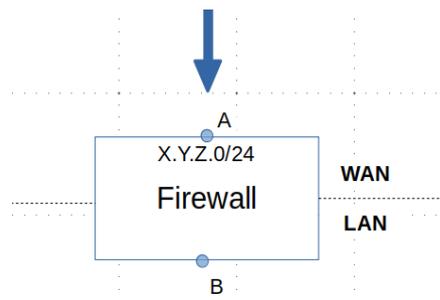

**FIGURE 3.** Logical representation of packet capture methodology for network interfaces on firewall appliances in the institutional network.

To conduct a comparative evaluation of the ML models, we also employed the UNSW-NB15 data set for benchmarking and validation, introduced below.

### 2) THE UNSW-NB15 DATA SET
The cybersecurity research group prepared the UNSW-NB15 data set at the Australian Center for Cyber Security (ACCS) [18], [21]. The simulated attacks may be grouped into nine classes, including fuzzers, analysis, backdoors, DoS, exploits, generic, reconnaissance, shellcode, and worms. The simulator updated itself with the information from a common vulnerabilities and exposures (CVE) website, and it created relevant attacks accordingly. For capturing the traffic, tcpdump was used. The framework architecture for the UNSW-NB15 data set generation is presented in Figure 4.

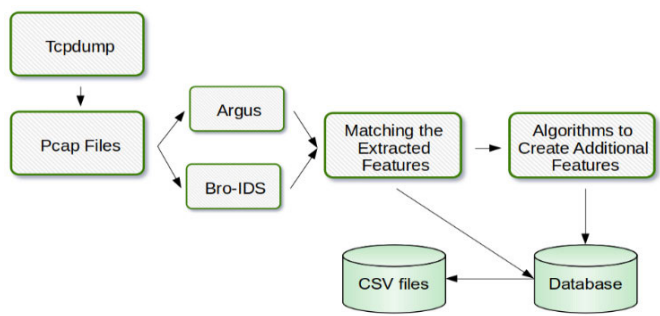

**FIGURE 4.** Framework architecture and workflow of generating the UNSW-NB15 data set.

The total data set is relatively large, with a total number of 2,500,000 flows/sessions. As it was done in the institutional data set, 100,000 flows were selected randomly. Below we present the pipeline for data processing.

This study was limited to the attack type of probing to keep the analyses manageable. It also allowed us to avoid accessing clear text application-layer payloads with sensitive PII (Personally Identifiable Information). The probing analysis is possible with the encrypted application layer payload. Utilizing the available information from transport and internet layers and header section from application layer (such as extracting clear text fields from HTTP header) provided a good feature set. Besides, metadata related to encrypted application layer payloads (e.g., size, flow duration) were also converted into features in this research. The identical approach was taken in the preparation of the UNSW-NB15 benchmarking data set. Originally it consists of nine different attack types, including the probing (cf. the reconnaissance phase) and a normal class. However, it was considered as a binary data set with probing and not probing classes only. Next, we describe the pipeline designed for data processing.

### B. THE PIPELINE
In the present study, we employed a supervised ML methodology for designing and developing an anomaly-based IDS. We trained the models for probing types of intrusion attempts rather than training the model for all attack types. Probing attacks consist of specific types of attacks, such as the *ping sweep*, which aims at discovering alive hosts, and *port scan*, which aims at discovering the services provided by the institution.

Probing constitutes the initial part of the reconnaissance phase of the cyber kill chain [22]. We chose *reconnaissance* from the UNSW-NB15 data set to accomplish benchmarking and validation, following the previous studies [23]. The packet captures were taken online. However, training and testing were performed offline due to practical limitations about lacking the authorization to apply tests or install new features on the production systems. The exact process was applied both on the institutional data set and the UNSW-NB15 data set. Since the data set preparation was done from scratch, it involves more steps than similar research done with public data sets. The preparation stage took most of the time course of the whole development (nearly 80%) reported by similar studies [24]. Figure 5 illustrates the data preparation pipeline.

We present the data collection and formatting processes and feature extraction, labeling, feature selection, model training, and testing below.

### 1) DATA COLLECTION AND FORMATTING
After packet captures were taken, they were divided into segments of 2 million packets to process the data conveniently. Then, network address translation (NAT) IP addresses were converted to public addresses. That was a necessary step for `argus`[2] and `tcptrace`[3] to extract flows from the packets

---
[2] https://openargus.org/
[3] http://www.tcptrace.org/

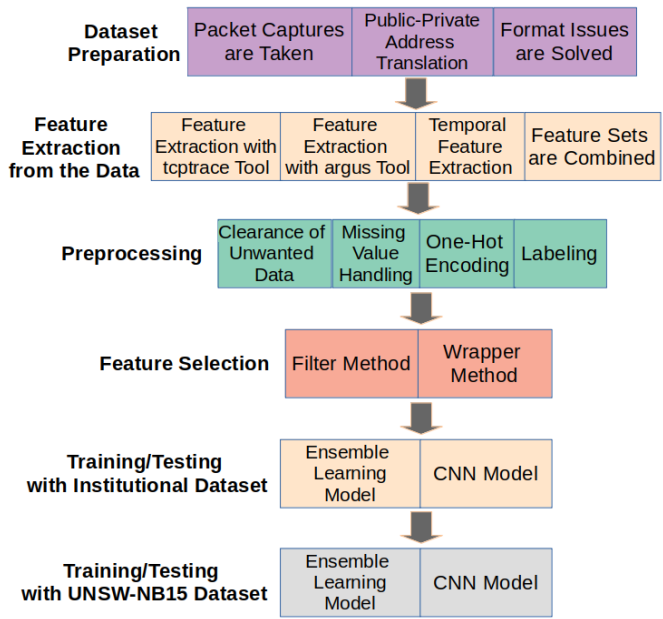

**FIGURE 5.** The pipeline for data preparation.

consistently with the IP addresses within a flow. We used the `tcprewrite` tool to overcome this problem.[4]

While capturing the traffic, the default format assigned by the underlying `libpcap` library of the `tcpdump` package was the Linux cooked capture (Linux_SLL) with no layer 2 (OSI model data link layer) address. To correctly analyze these packet captures by `wireshark`[5] and `argus` in later stages, pseudo-random MAC addresses were assigned by `tcprewrite`. We avoided a bias due to this arbitrary assignment since there was no feature extraction from layer 2 in the models. The next step in the pipeline was feature extraction, as presented below.

### 2) FEATURE EXTRACTION FROM THE INSTITUTIONAL DATA SET

Building a model based on flow-level intrusion detection is more common in IDS design since it has two significant advantages. First, more computational power and time are needed to build a packet-level IDS model. This is because a packet capture of 250 MBs includes around 2 million packets, whereas the same has only around 15,000 flows. Second, some attacks only become visible when packets are evaluated in their sessions. In some cases, the payload is only meaningful when it is complete after a session. We used `tcptrace` and `argus` for flow-level feature extraction from packet-level data (see [25], [26] for their documentation).

Custom features were then created for each flow. These were temporal features, also available in public data sets, such as NSL-KDD and UNSW-NB15. Those features are intended to catch anomalies by evaluating source behavior with time. `Pyshark` python library was used along with `pandas` and `numpy` libraries to develop the required scripts. Specific TCP header flags were considered as signs of probing behavior using nmap's scanning documentation. These were *ICMP*, *syn*, *syn-ack*, *null*, *fin*, *xmas* (*psh-urg-fin*), and *fin-ack*. In short, the scripts extracted the packets with one of the values in their TCP header. Then, evaluating by each flow, it counted the occurrence of each header value by source IP for 2 seconds. These values are expected to be an estimator for the source being malicious or normal. A sample output of temporal features is illustrated in Table 1.

[4]https://tcpreplay.appneta.com/wiki/tcprewrite
[5]https://www.wireshark.org

**TABLE 1.** Sample temporal features extracted with custom scripts from institutional packet captures.

| Start Time | SrcIP | SYN Count | ICMP Count | SYNACK Count | … |
|---|---|---|---|---|---|
| 156.93719 | 1.1.1.1 | 23 | 0 | 0 | ... |
| 157.50294 | 2.2.2.2 | 2 | 64 | 0 | ... |
| 157.80638 | 3.3.3.3 | 3 | 2 | 1 | … |
| … | … | … | … | … | … |

There were three sets of features present, from `tcptrace`, `argus`, and temporal features. Finally, these were combined, taking the session start time as the index column illustrated in Figure 6.

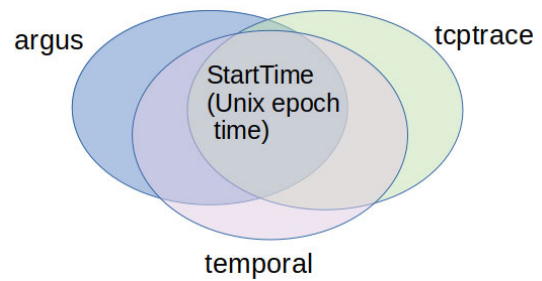

**FIGURE 6.** A combination of three sets of features was gathered from `argus` and `tcptrace` tools and the custom script based on the start time column.

A major limitation at this stage was that other flow types (UDP, ICMP) recorded by `argus` had their related columns as missing values for those two feature sets since `tcptrace` and temporal feature extraction script could extract features only from TCP sessions. This issue was addressed in the missing-value handling phase, as will be presented below.

### 3) PREPROCESSING AND LABELLING

The preprocessing step constitutes a critical phase in data analysis. At this stage, three main steps were executed with the help of `pandas` and `numpy` python libraries. First, redundant/unwanted columns were dropped. There were three types of unwanted features: The repeating columns between feature sets, the features with little or no variation, and the features with empty values. Second, categorical features were transformed into numerical ones, which is called one-hot encoding. Third, missing values were handled. While doing this, two distinct steps were taken. First, if a value was missing and it was probable for this sample to get a value, statistical summary functions, such as mean and median, were used to fill those missing values. For instance, a sample session could miss the "TotalBytes" feature due to a runtime problem. Because each session had a "TotalBytes" value, this could be replaced with the mean of the whole sample space. On the contrary, all UDP sessions missed the "TCPBaseAddress" feature, which was the intended behavior since UDP sessions do not have sequence numbers or base addresses. For such missing values, an impossible value was assigned as a placeholder.

After that, labeling was done. For this step, three sources were used. First, the firewall had a misuse-based IDS component, whose rules were live updated from the internet. Second, packet capture files were transferred into an environment with two open-source IDS software tools, namely `Snort`[6] and `Suricata`.[7] They were scanned with `Snort` and `Suricata`, and the created alerts were taken as labels. The test environment is illustrated in Figure 7. The rules were managed with `Pulled Pork`,[8] which downloads and installs rules per emerging threats.

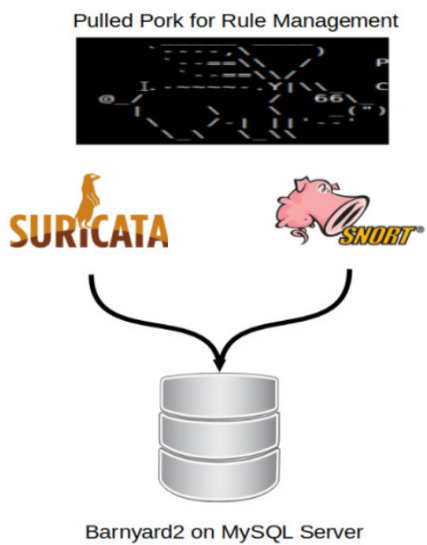

**FIGURE 7.** Test environment for labelling packet capture files taken from the institutional network.

After that, the `Barnyard2`[9] tool was used as a standard output medium of two systems. It provides a MySQL database environment for `Snort` and `Suricata` in which rules, alerts, configurations, and other metadata could be stored. Moreover, third, field expertise was used such that the authors know which target hosts are up or which services they are providing. After all, to make this process automated, it was included in a python script. The final labels were created by combining these three label-sets. The next step in the pipeline was feature selection, as presented below.

[6] https://www.snort.org/
[7] https://suricata-ids.org/
[8] https://github.com/shirkdog/pulledpork
[9] https://github.com/firnsy/barnyard2

### 4) FEATURE SELECTION

Feature selection is a critical step that affects the ML model performance directly. The previous studies have emphasized the significance of feature selection and proposed various methods for its execution [27]–[29]. Reducing the number of features has two main benefits in developing machine learning models. First, extracting a subset from the whole feature list helps to build a more accurate model by removing the noise caused by unnecessary features [29]. Second, model training time significantly decreases with a decreased number of features. The decrease is linear in the worst case depending on the training algorithm [30].

There are primarily two methods in feature selection. Filter methods measure each feature's correlation (independent variable) and the target (dependent) variable separately. The most prominent advantages of the filter method feature selection are that they are not computationally expensive and less prone to over-fitting [28]. On the other hand, the most significant drawback is that they cannot detect correlation and redundancy between the features. The second family of feature selection methods is the wrapper method. They have two different approaches; forward selection proposes bottom-up and backward elimination top-down strategy. In both ways, an ML model is built by selecting a learning algorithm. Based on its accuracy, features are either added or subtracted from the model until the best feature subset is generated. The most significant advantage of this method is that it provides a more accurate feature subset comparing to the filter method. However, there are disadvantages, including the requirement of high computational power and being prone to overfitting [30].

The proposed feature selection method in the present study aims at bringing together the strengths of the two approaches. For this, a threshold was identified, filter methods were applied, and the features were selected to satisfy those conditions. Then the wrapper method was used for further filtering. Figure 8 demonstrates the proposed workflow for the feature selection process. Such hybrid approaches have their frequent implementation on previous work and demonstrated efficiency [31], [32].

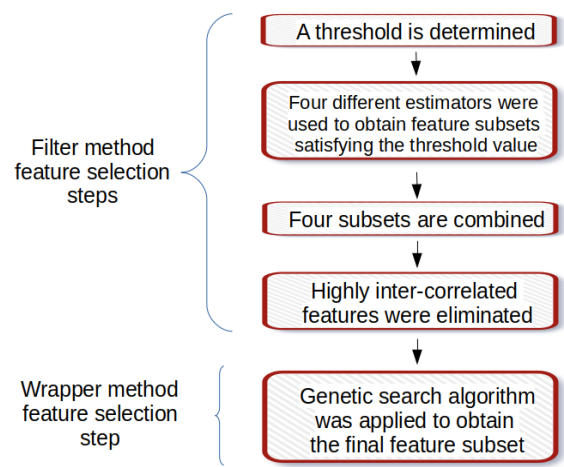

**FIGURE 8.** Proposed workflow for the feature selection process.

Numpy, pandas, sklearn, and genetic selection python libraries were utilized throughout the feature selection process. In the filter feature selection steps, four different estimators were used. In the first estimator, the features were selected if they correlated 0.3 or larger with the target variable. In the other three estimators, the best 25 features were selected using Chi-Square, ANOVA F-value, and extra-trees predictive methods, which returned mostly similar features. The four subsets were the combined, returning 32 features. Since the filter methods cannot detect the correlation between features, a correlation test was applied to eliminate ones that have more than 0.75 correlation among each other. After this test, 18 features remained (Table 12). An extensive description for each feature can be found in the manuals of the tools employed in the present study, such as [25], [26], [33].

The next phase in feature selection consisted of the wrapper method. Following the literature, the genetic search algorithm was employed [30], [31], [34], [35]. The algorithm was implemented by using python's sklearn-genetic library [36]. The random forest algorithm was used as the estimator, and the model was run through 50 generations. As a result, 13 features remained (Table 13).

The feature selection process was also applied to the UNSW-NB15 data set. The initial feature set consisted of 48 features. The number of features became 69 due to one-hot encoding in preprocessing. By the end of the preprocessing presented in Figure 8, 10 features were selected (Table 14 and 15).

### 5) MODEL TRAINING AND TESTING

This phase consisted of the development of two machine learning models. The models were selected based on the previous work on supervised ML models in designing anomaly-based IDSs [37]–[41]. In particular, an ensemble learning model was built upon the institutional data set, and a convolutional neural network (CNN) model followed that. After the evaluation of those two models, they were applied to the UNSW-NB15 data set for comparison.

Ensemble learning methods are meta-learners that utilize multiple ML models on the same data set to get more accurate results to reduce bias, noise, and variance. The adopted approach for ensemble training for the present study was the bagging method. Its alternatives (viz. the boosting method, the stacked method) are for a combination of weak learners that predict slightly better than random guessing, and they introduce a level of complexity to the model by adding one more layer. Moreover, bagging methods have the advantage of working in parallel, thus reducing the training time significantly. As the base classifier, we used naive Bayes, KNN, logistic regression, and SVM.

The second ML model was the CNN model. Besides being successful in image and video recognition, recommender systems, and natural language processing, CNN and its derivatives have their frequent implementation in the field of anomaly detection [39]–[41].

The two data sets (the institutional data set and the UNSW-NB15 data set) needed further preprocessing to fit the CNN model. Since the CNN needs the input in the form of 2D images, the vectors of 139 features (for each flow) were converted to 2D matrices by employing Naseer and Saleem's approach [42]. For this, 139 features were repeated seven times to form a $1 \times 973$ vector in which the $1^{st}$, the $140^{th}$, the $279^{th}$, ..., and the $835^{th}$ element were the same. Then, 51 more features were added as padding to get 1024 features which were later converted to $32 \times 32$ 2D features. An example flow, which was converted to a $32 \times 32$ grayscale image, is illustrated in Figure 9. The exact process was applied to the UNSW-NB15 data set, as well.

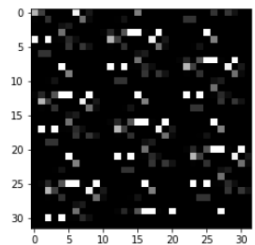

**FIGURE 9.** An example flow which was converted to 32 × 32 grayscale image.

There were two main issues in choosing the optimal size for the 2D image conversion. The first one was to obtain the best results following the predefined evaluation criteria. The second was to keep the model training, testing time, and data size manageable. Table 2 illustrates the alternative combinations. We selected an image size of $32 \times 32$ since it provided satisfying results in a manageable period.

**TABLE 2.** Summary of results from various image sizes for 2D conversion.

| Image Size | F1 Score | ROC AUC Score | Train/Test Duration for 10000 Samples | Data Size |
|---|---|---|---|---|
| 16 x 16 | 0.9818 | 0.9878 | 154 seconds | 20 MB |
| 32 x 32 | 0.9911 | 0.9927 | 343 seconds | 80 MB |
| 64 x 64 | 0.9929 | 0.9944 | 25 minutes | 330 MB |
| 128 x 128 | 0.9943 | 0.9958 | 114 minutes | 1.4 GB |

Hyperparameter tuning constitutes a crucial step in building any kind of neural network. There were hyperparameters such as batch size, number of epochs, filter size, optimization algorithm. Employing a trial-and-error method is not practical in such a scenario since the high number of hyperparameters generates too many combinations. To overcome this complexity, python's hyperas library, which works as an extension of the keras library, was employed. With the suggestions provided by hyperas, automated hyperparameter optimization was possible. As a result, the CNN model was built as depicted in Figure 10.

Another challenge in designing the ML model is bias. The most common bias is overfitting, which means that a model memorizes the training data and adapts itself too much to it.

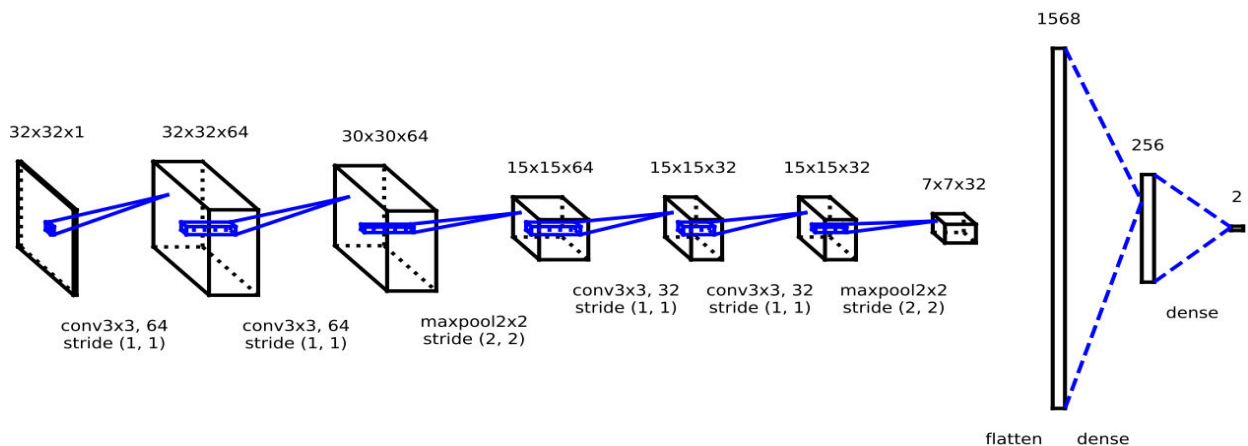

**FIGURE 10.** CNN model architecture for the institutional data set.

Overfitting has been a well-recognized problem [43], [44]. Solutions against overfitting include adding more data, generalizable data collection and inclusion to the model, building a simple model, and regularization [45]. In the present study, regularization was applied with dropout after each convolutional and fully connected (dense) layer.

The data sets were randomly divided into three sets (training, validation, and testing; 60%, 20%, and 20%, respectively) with no overlapping elements. The validation set was used for hyperparameter tuning. `Hyperopt` and its keras derivative `hyperas` python libraries were utilized for this. `Hyperopt` improves hyperparameter optimization by decreasing training time compared to the other standard grid search methods and random search [46]. Table 16-20 presents hyperparameters for both models. The models were then trained with those hyperparameters on training data. Finally, trained models were tested for accuracy. The results are presented in the following section.

## IV. RESULTS

In this section, the outcome of the models will be reported. The computer used for the development had the following specifications: i7-6700HQ 2.6 GHz CPU, NVIDIA GeForce GTX-950M GPU, 24 GBs of memory, and Linux Ubuntu 18.04 operating system. Python scripts were developed under various platforms, including PyCharm IDE, Anaconda Jupyter Notebook, and Google Colab.

Two criteria were employed as evaluation metrics: The F1 scores and the Receiver Operating Characteristics (ROC). The F1 scores consist of precision and recall values. The *precision* measures the ratio of correctly identified positive cases against all positive predicted cases, whereas the *recall* measures the ratio of correctly identified positive cases among all the actual positive cases. Based on these two metrics, the F1 score demonstrates the harmonic mean of precision and recall metrics and stands as a more accurate value for model evaluation. The Receiver Operating Characteristics (ROC) is widely accepted as a metric for ML models' performance assessment besides the F1 scores. It is used for assessing the relationship between recall (sensitivity) and precision (specificity) and depicts the model's performance for each possible threshold value for the classification. The Area Under Curve (AUC) constitutes the outcome of the ROC evaluation. A higher AUC means better performance of the model is evaluated. In the present study, the evaluation metrics were calculated with python `sklearn metrics` libraries throughout the implementation.

F1 and accuracy scores are frequently compared to determine the one which constitutes a better performance metric for an ML model. There are two justifications for the F1 score to be chosen instead of accuracy in this study. First, the rate of true negatives, which in most real-life scenarios are not significant as other classes in the confusion matrix, directly contributes to the accuracy score. However, the F1 score is only affected by true positives, false positives, and false negatives as equal to the harmonic mean of precision and recall values. Since this study aims to classify intrusion attempts and are labeled as positive in the models, the F1 score constitutes a more suitable performance evaluation metric. Second, the F1 score is more suitable with data sets with imbalanced distributions of positive and negative cases [47], [48]. More frequent classes manipulate the model in an imbalanced data set if the accuracy score is chosen as a performance evaluation metric since this value needs to be optimized. On the other hand, a less frequent class would be given less importance, and classification accuracy for the sample belonging to that class would decrease. This is important because positive instances represent the occurrences of attacks/intrusion attempts, usually constitute the minority of all cases in IDS data sets, and ignorance of them heavily distorts the model's reliability.

The second evaluation criterion was Receiver Operating Characteristics (ROC). It is widely accepted as a metric for ML models' performance assessment. It evaluates the relation between recall (sensitivity) and precision (specificity) and depicts the model's performance for every possible threshold value for classification. In other words, ROC is a dynamic evaluation of the model, whereas F1 may reflect the prediction capability in a specific threshold point. The area under the curve (AUC) constitutes the outcome of ROC evaluation, and a higher AUC means better performance of the model is evaluated. Apart from these two metrics, accuracy and false alarm rate are other frequently used metrics in anomaly-based intrusion detection research. Table 21 presents those metrics' results for both data sets and models. Evaluation metrics were calculated with python sklearn-metrics libraries throughout the implementation.

Table 3 illustrates the results for the ensemble models on the institutional data set. Table 4 illustrates the results for the UNSW-NB15 data set.

**TABLE 3.** Summary of results for institutional data set with ensemble learning model.

| Base Learner | F1 Score | ROC AUC Score |
|---|---|---|
| SVM | 0.9828 | 0.9832 |
| KNN | 0.9859 | 0.9862 |
| Naïve Bayes | 0.9767 | 0.9774 |
| Logistic Regression | 0.9813 | 0.9818 |

**TABLE 4.** Summary of results for UNSW-NB15 data set with ensemble learning model.

| Base Learner | F1 Score | ROC AUC Score |
|---|---|---|
| SVM | 0.9418 | 0.9859 |
| KNN | 0.9578 | 0.9778 |
| Naïve Bayes | 0.9198 | 0.9869 |
| Logistic Regression | 0.9584 | 0.9897 |

The model was trained and tested for five epochs until convergence. Table 5 shows the loss, accuracy, F1, and ROC AUC scores for the CNN model with the institutional data set. Table 6 presents the UNSW-NB15 data set results, where the model was trained and tested for seven epochs to reach convergence.

**TABLE 5.** Summary results of test data for the institutional data set with CNN model.

| Min Loss | Max Accuracy | F1 Score | ROC AUC Score |
|---|---|---|---|
| 0.0126 | 0.9952 | 0.9963 | 0.9998 |

**TABLE 6.** Summary results of test data for the UNSW-NB15 data set with CNN model.

| Min Loss | Max Accuracy | F1 Score | ROC AUC Score |
|---|---|---|---|
| 0.0211 | 0.9927 | 0.9902 | 0.9990 |

Feature visualization techniques were employed on institutional data to provide more insights into its behavior. t-SNE and saliency maps were two techniques, as shown in similar previous studies [49]–[51]. First, t-SNE was employed, which is a non-linear dimensionality reduction technique for reducing high-dimensional data into a lower dimension. Figure 11 represents a visualization of two classes with a randomly selected population of 25,000 flows. The visualization demonstrated that almost all flows appear in separate clusters. However, some overlap exists, and separation is not linearly possible. Second, saliency maps aids in visualizing the effect of every pixel in the input image to the predicted class. Figure 12 illustrates a random input's grayscale visualization and the corresponding saliency map. The guided method was chosen as the backpropagation modifier technique to obtain more refined results. The saliency map demonstrated the general alignment of input features and their importance. Expectedly, there are more and less significant pixel areas according to their contribution to the final prediction.

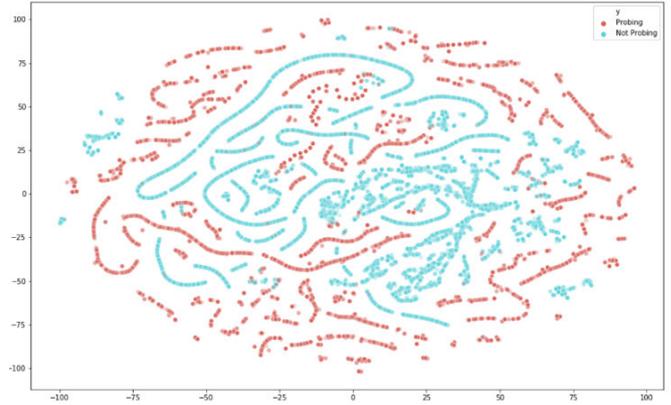

**FIGURE 11.** t-SNE visualization of probing and not probing classes within the institutional data set.

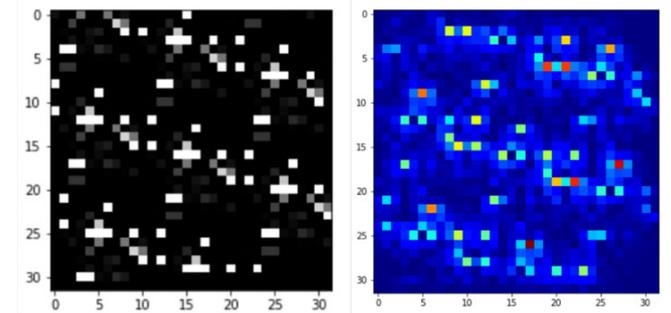

**FIGURE 12.** Grayscale visualization and saliency map of a randomly chosen flow from the institutional data set.

**TABLE 7.** Comparative summary results for both data sets and models.

| Data Set | Model | F1 Score | ROC AUC Score |
|---|---|---|---|
| Institutional | Ensemble with SVM | 0.9828 | 0.9832 |
| | Ensemble with KNN | 0.9859 | 0.9862 |
| | Ensemble with Naive Bayes | 0.9767 | 0.9774 |
| | Ensemble with Logistic Regression | 0.9813 | 0.9818 |
| | CNN | 0.9963 | 0.9998 |
| UNSW-NB15 | Ensemble with SVM | 0.9418 | 0.9859 |
| | Ensemble with KNN | 0.9578 | 0.9778 |
| | Ensemble with Naive Bayes | 0.9198 | 0.9869 |
| | Ensemble with Logistic Regression | 0.9584 | 0.9897 |
| | CNN | 0.9902 | 0.9990 |

**TABLE 8.** Hyperparameter optimization and training/testing duration of both models in institutional data set.

| Model | Environment | Hyperparameter Optimization | Model Training and Testing |
|---|---|---|---|
| Ensemble with SVM | | 36 minutes | 89 seconds |
| Ensemble with KNN | Local PC jupyter notebook | 15 minutes | 37 seconds |
| Ensemble with Naive Bayes | with CPU | 54 seconds | 48 seconds |
| Ensemble with Logistic Regression | | 4 minutes | 33 seconds |
| CNN | Google Colab with GPU | 66 minutes | 56 minutes |

In the following section, we present a discussion of the results with the framework of a comparative analysis of the two datasets and the previous findings in the literature.

## V. DISCUSSION

The results revealed several findings that need further evaluation. The results of the model tests are repeated in Table 7 for comparative analysis.

A major finding is that the CNN model outperformed the ensemble learning model in both data sets and for both evaluation metrics, suggesting that CNN models may have a higher potential for detecting network intrusions than ensemble models. For both models, the F1 scores from the institutional data set were slightly higher than those for the UNSW-NB15 data set. Finally, the ROC AUC scores were similar for both data sets and models. They were slightly higher in the UNSW-NB15 data set in three models of the ensemble classifier. So, one may say that the models returned similar results for the two data sets, indicating the analysis's reliability.

Despite the CNN model's higher performance compared to the ensemble model, it has drawbacks. Previous studies emphasize that some ML methods provide a more explainable structure about their inner mechanism while others do not [52]–[54]. The former category is called "white box," whereas the latter is "black box." As Riberio *et al.* argued, explainability is better evaluated on a scale than a binary classification [55], [56]. Methods such as decision trees, logistic regression, KNN, and SVM are placed closer to the white-box since it is possible to make inferences about their decision mechanisms. Vigano and Magazzeni argued that security is a critical domain in which less explainable ML models should be chosen with more caution [57].

Another drawback of employing a CNN model is its computational and temporal costs. Table 8 illustrates the hyperparameter optimization and model training times of both models for the institutional data set. The best-resulting ensemble model with the KNN base classifier needed 16 minutes for the whole process without GPU support. However, the CNN model running on the Google Colab Jupyter notebook environment with GPU support took 122 minutes for the same work. Although the models proposed in the present study were designed for running offline due to the limitations stated previously, employing a model with decreased training time would constitute a significant step in converting it to online.

**TABLE 9.** Results from recent studies based on UNSW-NB15 data set of results for UNSW-NB15 data set.

| Author(s) | F1 Score | ROC AUC Score | Accuracy | False Alarm Rate |
|---|---|---|---|---|
| Khammassi & Krichen [30] | 0.8287 | - | 0.8898 | 0.0706 |
| Souhail et. al. [60] | - | - | 0.8077 | - |
| Kumar et. al. [61] | 0.8655 | - | 0.9883 | 0.0095 |
| Karami [62] | 0.9183 | - | 0.9834 | 0.013 |
| Serkani et al. [63] | - | 0.99 | 0.9537 | 0.0203 |

**TABLE 10.** Misuse-based method confusion matrix on the institutional data set.

| | | True Condition | |
|---|---|---|---|
| | Total Population | Condition Positive | Condition Negative |
| **Predicted Condition** | Predicted Condition Positive | True Positive 48662 | False Positive 0 |
| | Predicted Condition Negative | False Negative 1944 | True Negative 49344 |

**TABLE 11.** Anomaly-based method (ensemble model with KNN base classifier) confusion matrix on the institutional data set.

| | | True Condition | |
|---|---|---|---|
| | Total Population | Condition Positive | Condition Negative |
| **Predicted Condition** | Predicted Condition Positive | True Positive 10133 | False Positive 137 |
| | Predicted Condition Negative | False Negative 152 | True Negative 9578 |

Table 9 demonstrates the results from recent studies based on the UNSW-NB15 data set for comparative evaluation. We approach the results presented in Table 8 cautiously since the presented studies trained one multi-class classifier to predict all ten classes of intrusion attempts and regular traffic. The presented scores belong to the reconnaissance type of intrusions for making comparisons. Moreover, each study utilized different ML models and compared them to get the best predictive model. So, the presented results are the best ones from the related studies. In particular, these results are from multi-class classifiers, and the previous studies demonstrate that it is not likely to obtain a binary classifier's performance with a vanilla multi-class classifier in each class's prediction [58], [59].

**TABLE 12.** Selected subset after filter method feature selection step for institutional data set.

| Number | Feature Name |
|---|---|
| 1 | mss_requested_a2b |
| 2 | DstTCPBase |
| 3 | min_segm_size_a2b |
| 4 | PCRatio |
| 5 | idletime_max_a2b |
| 6 | state_CON |
| 7 | state_REQ |
| 8 | state_RST |
| 9 | state_FIN |
| 10 | sTtl |
| 11 | dTtl |
| 12 | Dport |
| 13 | Dur |
| 14 | sMeanPktSz |
| 15 | dMeanPktSz |
| 16 | max_segm_size_a2b |
| 17 | FIN_pkts_a2b |
| 18 | adv_wind_scale_a2b |

**TABLE 13.** Final feature subset as the result of filter and wrapper method feature selection steps combined for institutional data set.

| Number | Feature Name |
|---|---|
| 1 | mss_requested_a2b |
| 2 | DstTCPBase |
| 3 | PCRatio |
| 4 | state_CON |
| 5 | state_RST |
| 6 | sTtl |
| 7 | dTtl |
| 8 | Dport |
| 9 | Dur |
| 10 | sMeanPktSz |
| 11 | dMeanPktSz |
| 12 | FIN_pkts_a2b |
| 13 | adv_wind_scale_a2b |

**TABLE 14.** Selected subset after filter method feature selection for UNSW-NB15 data set.

| Number | Feature Name |
|---|---|
| 1 | ct_dst_sport_ltm |
| 2 | ct_state_ttl |
| 3 | smeansz |
| 4 | dmeansz |
| 5 | ackdat |
| 6 | state_INT |
| 7 | Dload |
| 8 | service_dns |
| 9 | dsport |
| 10 | ct_ftp_cmd |
| 11 | dTtl |
| 12 | state_CON |
| 13 | proto_udp |
| 14 | service_- |
| 15 | Spkts |

**TABLE 15.** Final feature subset as the result of filter and wrapper method feature selection steps combined for UNSW-NB15 data set.

| Number | Feature Name |
|---|---|
| 1 | smeansz |
| 2 | dmeansz |
| 3 | ackdat |
| 4 | Dload |
| 5 | service_dns |
| 6 | dsport |
| 7 | dTtl |
| 8 | state_CON |
| 9 | service_- |
| 10 | Spkts |

The research question of the present study asked whether an anomaly-based ML model for IDS would have any success compared to the advantages offered by rule-based (misuse-based) intrusion detection methods, such as their simplicity. For this, we performed a further evaluation of the flows. Our findings showed that, among the randomly selected sample of 100,000 flows, 51,338 sessions were permitted by the firewall rules, and the remaining 48,662 sessions were dropped. However, out of those 51,338 permitted sessions, 1,994 were genuine probing attacks (as identified based on labeling methodology presented in Section 3). On the contrary, the anomaly-based model correctly classified a vast majority of those sessions. Table 10 and Table 11 present the individual performance of both methods.

The results show that the anomaly-based method predicted attack instances better than the misuse-based method since the recall score was higher in the former (98.52 % and 96.16%, respectively). Similarly, the F1 score of the anomaly-based method was higher than the latter, indicating a better overall performance (98.59% and 98.04%, respectively). Given that the total number of flows was 100,000, the corresponding number of attacks detected by one of the models but not the other may be critical for providing the organization's information security. To sum up, it can be argued that the anomaly-based ML model for IDS is more effective than the conventional misuse-based model.

**TABLE 16.** The optimized hyperparameters for ensemble learning model with SVM base learner for both data sets.

| Data Set | C | Degree | Kernel | Gamma | Max. Samples | Max Features | Bootstrap | Bootstrap Features |
|---|---|---|---|---|---|---|---|---|
| **Institutional** | 47 | 4.25 | RBF | 1.64 | 0.7514 | 0.8762 | True | False |
| **UNSW-NB15** | 4.59 | 5.15 | Poly | 1.67 | 0.9170 | 0.9642 | False | False |

**TABLE 17.** The optimized hyperparameters for ensemble learning model with KNN base learner for both data sets.

| Data Set | Algorithm | Leaf Size | Number of Neighbours | p | Weights | Max. Samples | Max Features | Bootstrap | Bootstrap Features |
|---|---|---|---|---|---|---|---|---|---|
| **Institutional** | Auto | 26 | 3 | 1 | Distance | 0.9685 | 0.9986 | True | False |
| **UNSW-NB15** | Brute | 24 | 3 | 1 | Distance | 0.8422 | 0.9583 | True | True |

**TABLE 18.** The optimized hyperparameters for ensemble learning model with naive bayes base learner for both data sets.

| Data Set | Variance Smoothing | Max. Samples | Max Features | Bootstrap | Bootstrap Features |
|---|---|---|---|---|---|
| **Institutional** | 3.15e-05 | 0.4281 | 0.7954 | False | True |
| **UNSW-NB15** | 0.28 | 0.5793 | 0.9409 | True | False |

**TABLE 19.** The optimized hyperparameters for ensemble learning model with logistic regression base learner for both data sets.

| Data Set | C | Max. Iterations | Multi Class | Penalty | Solver | Tolerance | Max. Samples | Max Features | Bootstrap | Bootstrap Features |
|---|---|---|---|---|---|---|---|---|---|---|
| **Institutional** | 190 | 200 | Auto | None | LBFGS | 0.0673 | 0.0692 | 0.8906 | False | False |
| **UNSW-NB15** | 80 | 220 | Multinomial | None | LBFGS | 0.6888 | 0.8170 | 0.9098 | True | False |

**TABLE 20.** The optimized hyperparameters for CNN model for both data sets.

| Data Set | Number of Conv. Layers | Kernel Size | Activation Function | Dropout | Loss Function | Optimizer | Batch Size | Number of Epochs |
|---|---|---|---|---|---|---|---|---|
| **Institutional** | 3 | 64,64,64 | Sigmoid, ReLU, Sigmoid | 0.12,0.16,0.11 | Categorical Crossentropy | Adam | 128 | 5 |
| **UNSW-NB15** | 4 | 64,64,32,64 | ReLU,Sigmoid, ReLU,ReLU | 0.54,0.43,0.69 | Categorical Crossentropy | Rmsprop | 128 | 7 |

## VI. CONCLUSION AND FUTURE WORK

The present study investigated the potential of an anomaly-based ML model for IDS compared to the misuse-based models. An institutional data set was utilized to reveal whether a model could be designed that satisfies validity and provide better accuracy. We tested the validity of the approach by designing two ML models. The models were then run on the institutional data set and the UNSW-NB15 data set for benchmarking. To sum things up, beforementioned findings demonstrated three points. First, the results showed that the methodology was robust enough to provide the validity of the study. Approximate outcomes for both institutional and UNSW-NB15 data sets justified the proposed approach.

Second, two ML models were trained to compare their performance. Both of them returned satisfactory ROC AUC and F1 scores for the chosen attack type, namely probing. The CNN model did slightly better than the ensemble models on the institutional data set. However, the ensemble model is advantageous against the CNN model given better 'explainability' and lower demand for computational resources than the CNN model.

Third, the proposed anomaly-based IDS models obtained better results compared to the actual misuse based classification within the real life institutional environment. Unlike the ubiquitous work previously done with simulated data sets, this finding is novel and encouraged the future use of anomaly-based techniques in enterprise IDS.

TABLE 21. Accuracy and false alarm rate scores for both data sets and models.

| Data Set | Model | Accuracy | False Alarm Rate |
|---|---|---|---|
| Institutional | Ensemble Learning with SVM | 0.9768 | 0.0010 |
| | Ensemble Learning with KNN | 0.9931 | 0.0010 |
| | Ensemble Learning with Naive Bayes | 0.9806 | 0.0011 |
| | Ensemble Learning with Logistic Regression | 0.9790 | 0.0010 |
| | CNN | 0.9968 | 0.0008 |
| UNSW-NB15 | Ensemble Learning with SVM | 0.9902 | 0.0051 |
| | Ensemble Learning with KNN | 0.9960 | 0.0051 |
| | Ensemble Learning with Naive Bayes | 0.9818 | 0.0049 |
| | Ensemble Learning with Logistic Regression | 0.9896 | 0.0037 |
| | CNN | 0.9885 | 0.0041 |

Future research is necessary to address the limitations of the present study. First, in its recent form, the proposed pipeline has been designed to work offline. Moving the pipeline to an online platform may result in novel requirements that have not been foreseen in the present study. Second, the present study focused on a specific and frequently seen type of attack, namely probing. Accordingly, the ML models were binary classifiers that classify each flow either as a probing attack or not. Future research should aim at expanding the proposed methodology for other attack types. Two possible approaches could be taken for that. First, a multi-class classifier for classification of different attack categories at once. Second, multiple one-vs-all binary classifiers that run in parallel to identify each class of intrusion attempt separately.

Another limitation of the study was that the institutional data set was gathered from a specific production environment. Although the data set had the advantage of including actual cyber attacks from the outside of the organization, its coverage is limited to the attack types related to the organization's services. Future research should also address improving the missing-data handling methodology since that is a significant challenge for designing heterogeneous data sets. Finally, as recognized in the relevant work, automated labeling is a common pitfall in supervised learning in IDS models, so it needs improvement since it is prone to error in its recent form.

## APPENDIX
See Tables 12–21.

## REFERENCES

[1] *CybersecurityUpdate—WebProNews*. Accessed: Aug. 21, 2020. [Online]. Available: https://www.webpronews.com/cisco-cybersecurity-threats/
[2] Z. M. Smith, E. Lostri, and J. A. Lewis, "The hidden costs of cybercrime," in *Proc. McAfee*, 2020, p. 3.
[3] *Cybersecurity Report 2020*, Check Point Research, 2020, p. 35.
[4] C. Douligeris, "From January 2019 to April 2020 the year in review ENISA threat Landscap," ENISA, Tech. Rep., 2020.
[5] D. Singh and V. P. Singh, "Comparative study of various distributed intrusion detection systems for WLAN," *Global J. Res. Eng. Electr. Electron. Eng.*, vol. 12, no. 6, pp. 49–56, 2012.
[6] K. Rahul-Vigneswaran, P. Prabaharan, and K. P. Soman, "A compendium on network and host based intrusion detection systems," in *Proc. Int. Conf. Data Sci., Mach. Learn. Appl.*, Singapore, 2019, pp. 23–30.
[7] M. Ring, S. Wunderlich, D. Scheuring, D. Landes, and A. Hotho, "A survey of network-based intrusion detection data sets," *Comput. Secur.*, vol. 86, pp. 147–167, Sep. 2019.
[8] A. L. Buczak and E. Guven, "A survey of data mining and machine learning methods for cyber security intrusion detection," *IEEE Commun. Surveys Tuts.*, vol. 18, no. 2, pp. 1153–1176, 2nd Quart., 2016.
[9] D. K. Bhattacharyya and J. K. Kalita, *Network Anomaly Detection: A Machine Learning Perspective*. Boca Raton, FL, USA: CRC Press, 2013.
[10] N. J. Nilsson, "Artificial intelligence: A modern approach," *Artif. Intell.*, vol. 82, nos. 1–2, pp. 369–380, Apr. 1996.
[11] R. Chitrakar and H. Chuanhe, "Anomaly detection using support vector machine classification with k-medoids clustering," in *Proc. 3rd Asian Himalayas Int. Conf. Internet*, Nov. 2012, pp. 1–5.
[12] I. P.-B. A. Syarif and G. Wills, "Unsupervised clustering approach for network anomaly detection," in *Proc. Int. Conf. Netw. Digit. Technol.*, Berlin, Germany, 2012, pp. 135–145.
[13] K. Moh, M. Aung, and N. N. Oo, "Association rule pattern mining approaches network anomaly detection," in *Proc. Int. Conf. Future Comput. Technol.*, Singapore, 2015, pp. 164–170.
[14] A. H. Hamamoto, L. F. Carvalho, L. D. H. Sampaio, T. Abrão, and M. L. Proença, "Network anomaly detection system using genetic algorithm and fuzzy logic," *Expert Syst. Appl.*, vol. 92, pp. 390–402, Feb. 2018.
[15] N. T. Pham, E. Foo, S. Suriadi, H. Jeffrey, and H. F. M. Lahza, "Improving performance of intrusion detection system using ensemble methods and feature selection," in *Proc. Australas. Comput. Sci. Week Multiconference*, Jan. 2018, pp. 1–6.
[16] I. Sharafaldin, A. Gharib, A. H. Lashkari, and A. A. Ghorbani, "Towards a reliable intrusion detection benchmark dataset," *Softw. Netw.*, vol. 2017, no. 1, pp. 177–200, 2017.
[17] A. M. Al Tobi and I. Duncan, "KDD 1999 generation faults: A review and analysis," *J. Cyber Secur. Technol.*, vol. 2, nos. 3–4, pp. 164–200, Oct. 2018.
[18] N. Moustafa and J. Slay, "UNSW-NB15: A comprehensive data set for network intrusion detection systems," in *Proc. Mil. Commun. Inf. Syst.*, 2015, pp. 1–6.
[19] I. Sharafaldin, A. Habibi Lashkari, and A. A. Ghorbani, "Toward generating a new intrusion detection dataset and intrusion traffic characterization," in *Proc. 4th Int. Conf. Inf. Syst. Secur. Privacy*, 2018, pp. 108–116.
[20] *Hulk—Packet Storm*. Accessed: Aug. 22, 2020. [Online]. Available: https://packetstormsecurity.com/files/112856/HULK-Http-Unbearable-Load-King.html
[21] N. Moustafa and J. Slay, "The evaluation of network anomaly detection systems: Statistical analysis of the UNSW-NB15 data set and the comparison with the KDD99 data set," *Inf. Secur. J., Global Perspective*, vol. 25, nos. 1–3, pp. 18–31, Apr. 2016.
[22] *Cyber Kill Chain—Lockheed Martin*. Accessed: Aug. 27, 2020. [Online]. Available: https://www.lockheedmartin.com/en-us/capabilities/cyber-kill-chain.html
[23] A. Divekar, M. Parekh, V. Savla, R. Mishra, and M. Shirole, "Benchmarking datasets for anomaly-based network intrusion detection: KDD CUP 99 alternatives," in *Proc. IEEE 3rd Int. Conf. Comput., Commun. Secur. (ICCCS)*, Kathmandu, Nepal, Oct. 2018, pp. 1–8.
[24] P. Gil. *Cleaning Big Data—Forbes*. Accessed: Aug. 26, 2020. [Online]. Available: https://www.forbes.com/sites/gilpress/2016/03/23/data-preparation-most-time-consuming-least-enjoyable-data-science-task-survey-says/#79e15eaa6f63
[25] *Documentation—Argus* Accessed: Aug. 27, 2020. [Online]. Available: https://openargus.org/documentation,
[26] *Online Manual—Tcptrace*. Accessed: Aug. 27, 2020. [Online]. Available: http://www.tcptrace.org/manual.html
[27] M. Alkasassbeh, "An empirical evaluation for the intrusion detection features based on machine learning and feature selection methods," *J. Theor. Appl. Inf. Technol.*, vol. 95, no. 22, pp. 5962–5976, 2017.
[28] M. A. Ambusaidi, X. He, P. Nanda, and Z. Tan, "Building an intrusion detection system using a filter-based feature selection algorithm," *IEEE Trans. Comput.*, vol. 65, no. 10, pp. 2986–2998, Oct. 2016.

[29] V. Bolón-Canedo, N. Sánchez-Maroño, and A. Alonso-Betanzos, "Feature selection for high-dimensional data," *Prog. Artif. Intell.*, vol. 5, no. 2, pp. 65–75, 2016.

[30] C. Khammassi and S. Krichen, "A GA-LR wrapper approach for feature selection in network intrusion detection," *Comput. Secur.*, vol. 70, pp. 255–277, Sep. 2017.

[31] M. A. Ambusaidi, X. He, Z. Tan, P. Nanda, L. F. Lu, and U. T. Nagar, "A novel feature selection approach for intrusion detection data classification," in *Proc. IEEE 13th Int. Conf. Trust, Secur. Privacy Comput. Commun.*, Beijing, China, Sep. 2014, pp. 82–89.

[32] S. Mohammadi, H. Mirvaziri, M. Ghazizadeh-Ahsaee, and H. Karimipour, "Cyber intrusion detection by combined feature selection algorithm," *J. Inf. Secur. Appl.*, vol. 44, pp. 80–88, Feb. 2019.

[33] *Discriminators for use in flow-based classification*, Queen Mary University of London, London, U.K., 2005, pp. 1–14.

[34] *Introduction to Genetic Algorithms—Towards Data Science*. Accessed: Aug. 27, 2020. [Online]. Available: https://towardsdatascience.com/introduction-to-genetic-algorithms-including-example-code-e396e98d8bf3

[35] *Feature Selection using Genetic Algorithms in R—Towards Data Science*. Accessed: Aug. 26, 2020. [Online]. Available: https://towardsdatascience.com/feature-selection-using-genetic-algorithms-in-r-3d9252f1aa66

[36] *sklearn-Genetic—GitHub*. Accessed: Aug. 27, 2020. [Online]. Available: https://github.com/manuel-calzolari/sklearn-genetic

[37] D. P. Gaikwad and R. C. Thool, "Intrusion detection system using bagging ensemble method of machine learning," in *Proc. Int. Conf. Comput. Commun. Control Autom.*, Maharashtra, India, Feb. 2015, pp. 291–295.

[38] J. Vanerio and P. Casas, "Ensemble-learning approaches for network security and anomaly detection," in *Proc. Workshop Big Data Anal. Mach. Learn. Data Commun. Netw.*, Los Angeles, CA, USA, 2017, pp. 1–6.

[39] M. Kravchik and A. Shabtai, "Detecting cyber attacks in industrial control systems using convolutional neural networks," in *Proc. ACM Conf. Comput. Commun. Secur.*, Toronto, ON, Canada, 2018, pp. 72–83.

[40] D. Kwon, K. Natarajan, S. C. Suh, and H. Kim, "An empirical study on network anomaly detection using convolutional neural networks," in *Proc. Int. Conf. Distrib. Comput. Syst.*, Vienna, Austria, 2018, pp. 1595–1598.

[41] M. Zhu, K. Ye, Y. Wang, and C. Z. Xu, "A deep learning approach for network anomaly detection based on AMF-LSTM," in *Proc. IFIP Int. Conf. Netw. Parallel Comput., Muroran, Jpn.*, 2018, pp. 137–141.

[42] S. Naseer and Y. Saleem, "Enhanced network intrusion detection using deep convolutional neural networks," *KSII Trans. Internet Inf. Syst.*, vol. 12, no. 10, pp. 5159–5178, 2018.

[43] T. Dietterich, "Overfitting and undercomputing in machine learning," *ACM Comput. Surv.*, vol. 27, no. 3, pp. 326–327, Sep. 1995.

[44] C. Schaffer, "Overfitting avoidance as bias," *Mach. Learn.*, vol. 10, no. 2, pp. 153–178, Feb. 1993.

[45] *Deep Learning #3: More on CNNs & Handling Overfitting—Towards Data Science*. Accessed: Aug. 27, 2020. [Online]. Available: https://towardsdatascience.com/deep-learning-3-more-on-cnns-handling-overfitting-2bd5d99abe5d

[46] J. Bergstra, D. Yamins, and D. Cox, "Hyperopt: A Python library for optimizing the hyperparameters of machine learning algorithms," in *Proc. 12th Python Sci. Conf.*, Austin, TX, USA, 2013, pp. 1–8.

[47] J. O. Berger, *Statistical Decision Theory and Bayesian Analysis*. Springer, 2013.

[48] P. Powers, "Evaluation: From precision, recall and F-measure to ROC, informedness, markedness and correlation," *J. Mach. Learn. Technol.*, vol. 2, pp. 37–63, Feb. 2011.

[49] R. Vinayakumar, M. Alazab, K. P. Soman, P. Poornachandran, A. Al-Nemrat, and S. Venkatraman, "Deep learning approach for intelligent intrusion detection system," *IEEE Access*, vol. 7, pp. 41525–41550, 2019.

[50] S. Sriram, R. Vinayakumar, M. Alazab, and S. Kp, "Network flow based IoT botnet attack detection using deep learning," in *Proc. IEEE Conf. Comput. Commun. Workshops (INFOCOM WKSHPS)*, Toronto, ON, Canada, Jul. 2020, pp. 189–194.

[51] R. Vinayakumar, K. P. Soman, and P. Poornachandran, "Applying convolutional neural network for network intrusion detection," in *Proc. Int. Conf. Adv. Comput., Commun. Inform.*, Udupi, India, 2017, pp. 1222–1228.

[52] T. Miller, "Explanation in artificial intelligence: Insights from the social sciences," *Artif. Intell.*, vol. 267, pp. 1–39, Oct. 2019.

[53] A. Barredo Arrieta, N. Díaz-Rodríguez, J. Del Ser, A. Bennetot, S. Tabik, A. Barbado, S. Garcia, S. Gil-Lopez, D. Molina, R. Benjamins, R. Chatila, and F. Herrera, "Explainable artificial intelligence (XAI): Concepts, taxonomies, opportunities and challenges toward responsible AI," *Inf. Fusion*, vol. 58, pp. 82–115, Jun. 2020.

[54] F. K. Došilović, M. Brčić, and N. Hlupić, "Explainable artificial intelligence: A survey," in *Proc. Int. Conv. Inf. Commun. Technol., Electron. Microelectron.*, Opatija, Croatia, 2018, pp. 210–215.

[55] M. T. Ribeiro, S. Singh, and C. Guestrin, "'Why should I trust you?' Explaining the predictions of any classifier," in *Proc. SIGKDD Int. Conf. Knowl. Discovery Data Mining*, London, U.K., 2016, pp. 1135–1144.

[56] A. Adadi and B. M., "Peeking inside the black-box: A survey on explainable artificial intelligence," *IEEE Access*, vol. 6, pp. 52138–52160, 2018.

[57] *Explainable Security*. Accessed: Aug. 27, 2020. [Online]. Available: https://arxiv.org/pdf/1807.04178.pdf

[58] D. Silva-Palacios, C. Ferri, and M. J. Ramírez-Quintana, "Improving performance of multiclass classification by inducing class hierarchies," *Procedia Comput. Sci.*, vol. 108, pp. 1692–1701, Dec. 2017.

[59] B. Krawczyk, M. Galar, M. Woániak, H. Bustince, and F. Herrera, "Dynamic ensemble selection for multi-class classification with one-class classifiers," *Pattern Recognit.*, vol. 83, pp. 34–51, Nov. 2018.

[60] S. Meftah, T. Rachidi, and N. Assem, "Network based intrusion detection using the UNSW-NB15 dataset," *Int. J. Comput. Digit. Syst.*, vol. 8, no. 5, pp. 478–487, 2019.

[61] V. Kumar, D. Sinha, A. K. Das, S. C. Pandey, and R. T. Goswami, "An integrated rule based intrusion detection system: Analysis on UNSW-NB15 data set and the real time online dataset," *Cluster Comput.*, vol. 23, no. 2, pp. 1397–1418, Jun. 2020.

[62] A. Karami, "An anomaly-based intrusion detection system in presence of benign outliers with visualization capabilities," *Expert Syst. Appl.*, vol. 108, pp. 36–60, Oct. 2018.

[63] E. Serkani, G. H. Gharaee, and N. Mohammadzadeh, "Anomaly detection using SVM as classifier and decision tree for optimizing feature vectors," *Int. J. Inf. Secur.*, vol. 11, no. 2, pp. 159–171, 2019.

○ ○ ○